\documentclass[
aps,prl,reprint,superscriptaddress,
]{revtex4-2}

\usepackage{graphicx}
\usepackage{dcolumn}
\usepackage{bm}
\usepackage{etoolbox}
\usepackage{color}
\usepackage{xcolor}
\usepackage{amsmath} 
\usepackage{enumitem}

\begin{document}


\title{Resonance density range governs two-plasmon decay saturation and enables hot-electron prediction in inertial confinement fusion
}

\author{C. Yao}
\affiliation{ 
Department of Plasma Physics and Fusion Engineering and CAS Key Laboratory of Frontier Physics in Controlled Nuclear Fusion, University of Science and Technology of China, Hefei, Anhui 230026, People’s Republic of China
}%
\affiliation{ 
Institute of Applied Physics and Computational Mathematics, Beijing 100088, People’s Republic of
China
}%

\author{J. Li}%
\thanks{Corresponding author: junlisu@ustc.edu.cn}
 \affiliation{ 
Department of Plasma Physics and Fusion Engineering and CAS Key Laboratory of Frontier Physics in Controlled Nuclear Fusion, University of Science and Technology of China, Hefei, Anhui 230026, People’s Republic of China
}%
\affiliation{ 
Collaborative Innovation Center of IFSA (CICIFSA), Shanghai Jiao Tong University, Shanghai 200240, People’s Republic of China
}%

\author{L. Hao}
\thanks{Corresponding author: hao\_liang@iapcm.ac.cn}
\affiliation{%
Institute of Applied Physics and Computational Mathematics, Beijing 100088, People’s Republic of
China
}%

\author{R. Yan}%
\thanks{Corresponding author: ruiyan@ustc.edu.cn}
\affiliation{ 
Department of Modern Mechanics, University of Science and Technology of China, Hefei, Anhui
230026, People’s Republic of China
}%
\affiliation{ 
Collaborative Innovation Center of IFSA (CICIFSA), Shanghai Jiao Tong University, Shanghai 200240, People’s Republic of China
}%

\author{T.Tao}%
 \affiliation{ 
Department of Plasma Physics and Fusion Engineering and CAS Key Laboratory of Frontier Physics in Controlled Nuclear Fusion, University of Science and Technology of China, Hefei, Anhui 230026, People’s Republic of China
}%

\author{G-N.Zheng}%
 \affiliation{ 
Department of Plasma Physics and Fusion Engineering and CAS Key Laboratory of Frontier Physics in Controlled Nuclear Fusion, University of Science and Technology of China, Hefei, Anhui 230026, People’s Republic of China
}%

\author{Q.Jia}%
 \affiliation{ 
Department of Plasma Physics and Fusion Engineering and CAS Key Laboratory of Frontier Physics in Controlled Nuclear Fusion, University of Science and Technology of China, Hefei, Anhui 230026, People’s Republic of China
}%

\author{Y-K. Ding}
\affiliation{%
Institute of Applied Physics and Computational Mathematics, Beijing 100088, People’s Republic of
China
}%

\author{J. Zheng}%
 \affiliation{ 
Department of Plasma Physics and Fusion Engineering and CAS Key Laboratory of Frontier Physics in Controlled Nuclear Fusion, University of Science and Technology of China, Hefei, Anhui 230026, People’s Republic of China
}%
\affiliation{ 
Collaborative Innovation Center of IFSA (CICIFSA), Shanghai Jiao Tong University, Shanghai 200240, People’s Republic of China
}%

\date{\today}

\begin{abstract}



The saturation level of parametric instabilities critically determines their impact on fusion plasmas. 
We identify the resonance density range of two-plasmon decay as the critical parameter governing nonlinear saturation of ion density fluctuations and Langmuir waves, which drive hot-electron generation. 
Using this insight, we develop a predictive scaling model for the hot-electron energy fraction $f_{hot}$ that depends only on the laser intensity $I$, with plasma conditions encoded via plasma ablation theory. 
The model can work for various experimental configurations—requiring only two ($I$, $f_{hot}$) data points to calibrate coefficients—and successfully reproduces results from prior OMEGA and OMEGA-EP experiments. 

\end{abstract}

\maketitle


Two-plasmon decay (TPD) instability, a parametric process in which an electromagnetic pump wave decays into two plasmons \cite{Kruer2003}, poses a critical challenge for fusion energy research. \textcolor{black}{In inertial confinement fusion (ICF), TPD drives anomalous laser energy absorption \cite{Li2017a,Turnbull2020} and generates hot electrons \cite{Hohenberger2015,Michel2012a} that preheat the deuterium–tritium fuel, thus severely degrading implosion performance \cite{Christopherson2021,Solodov2022}}. TPD has been identified as a principal source of hot electrons in numerous direct-drive ICF experiments on OMEGA laser facility \cite{Seka2009a,Michel2012a} and National Ignition Facility (NIF) \cite{Hohenberger2015}. In indirect-drive ICF, TPD has also been implicated in the generation of detrimental hot electrons \cite{Regan2010}. \textcolor{black}{Besides, in magnetic confinement fusion, TPD impairs electron cyclotron resonance heating efficiency \cite{Clod2024,Tancetti2022,YuPopov2015}}.These observations underscore the critical importance of mitigating TPD in pursuit of successful high-gain fusion energy applications.

In ICF, the detrimental impact of TPD-generated hot electrons is primarily governed by the amplitudes of daughter electron plasma waves (EPWs) at nonlinear saturation. The preceding linear growth stage—during which only absolute TPD modes grow appreciably—persists for merely a few picoseconds before nonlinear saturation occurs, a duration much shorter than the overall laser pulse timescale ($\sim$ns). \textcolor{black}{In the subsequent nonlinear stage, prior studies have demonstrated that hot electron generation is predominantly governed by a staged acceleration mechanism driven by both convective and absolute TPD modes~\cite{Yan2012,Yan2009}, which exhibit comparable saturation amplitudes.}
However, predicting the amplitudes and spectra of these modes is highly challenging due to the involvement of complex physical processes, including Langmuir decay instability (LDI)~\cite{DuBois1995,Wen2019}, ion density fluctuations~\cite{Langdon1979}, convective-to-absolute transitions~\cite{Li2017b}, and interactions among different laser-plasma instabilities (LPIs)~\cite{Cao2020}. As a result, earlier studies~\cite{Michel2012a,Cao2022} have bypassed direct evaluation of EPW amplitudes and instead focused on developing empirical scalings for the hot electron energy fraction, $f_{\mathrm{hot}}$, based on experimental or numerical data. While this approach yields practical estimates of $f_{\mathrm{hot}}$ for specific experimental conditions, it typically requires a large volume of experimental data under similar configurations to improve predictive reliability.
Without sufficient understanding of EPW amplitudes, the predictive reliability may deteriorate when applied to experimental configurations or physical parameters beyond those directly constrained by existing data.

In this Letter, we investigate $f_{hot}$ based on a novel physical understanding of the saturation amplitudes of TPD-driven EPWs. 
We uncover new insights into absolute TPD in inhomogeneous plasmas by identifying the resonance density range, which can be directly derived from the standard TPD theory, as the key parameter governing the nonlinear saturation of TPD.
We demonstrate that this density range is physically aligned with the ion density fluctuation level at TPD saturation, allowing straightforward calculation of the saturated electric field. 
Building on this insight, we construct a predictive scaling model for $f_{\mathrm{hot}}$ that depends solely on the laser intensity $I$, with plasma conditions incorporated via plasma ablation theory. The model applies across a range of experimental conditions, requiring only two calibration points—$(I, f_{\mathrm{hot}})$ pairs—to determine coefficients for a specific configuration. 
It successfully reproduces results from prior OMEGA and OMEGA-EP experiments, providing a compact, experimentally anchored framework for predicting hot-electron generation.

\begin{figure}[h]
\includegraphics[width=1.0\linewidth]{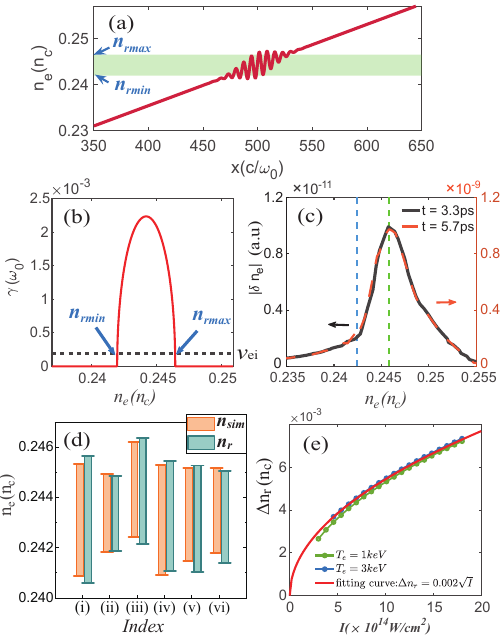}%
\caption{\label{fig:1} 
\textcolor{black}{(a) Schematic diagram of the resonance density range $n_r$ and ion density fluctuations at TPD saturation.}
(b) 
Linear growth rates of the TPD mode $\vec{k}_a = (k_{ax} = 0.88 \omega_0/c, k_{ay} = 0.091 \omega_0/c)$ at different densities for $T_e = 2.5$~keV, $I = 6 \times  10^{14}W/cm^2$,$L_n = 150\mu m$. The dashed horizontal line represents $\nu_{ei}$.
(c)–(d) LTS fluid simulations results across parameter space $(T_e,L_n,I_0)$ (unit: $keV,\mu m,10^{14}W/cm^2$): (i) (3.0, 100, 8.5), (ii) (3.0, 300, 3.0), (iii) (2.5, 150, 6.0), (iv) (3.0, 150, 6.5), (v) (3.0, 150, 6.0), (vi) (3.0, 200, 4.5). 
(c)
Spatial envelopes of $|\delta n_e|$ at 3.3 ps (black) and 5.7ps (orange) from simulation (iii).  Vertical lines mark the locations with maximum (blue) and minumum (green) second-order derivatives at $n_e = 0.2424n_c$ and $0.2462n_c$, respectively.
(d) Theoretical $n_{r}$ (green) and the simulated resonance density ranges $n_{sim}$ (orange). (e)
Theoretical relation between $\Delta n_{r}$ and $I$ for $T_e=$ 1 and 3 $keV$ with fitted curve.}
\end{figure}

\textcolor{black}{We first highlight the critical role of ion density fluctuations ($\Delta n_i$) in TPD saturation. Previous studies have shown that $\Delta n_i$ and TPD wave modes exhibit either quasi-steady states~\cite{Yan2012,Zhang2014,Li2017a} or recurrent behaviors with opposite phases~\cite{Yan2009,Langdon1979}, indicating that saturation occurs when $\Delta n_i$ becomes sufficiently large to suppress TPD growth. Below, we demonstrate that the saturation level of $\Delta n_i$ is closely linked to a newly defined resonance density range, $n_r = (n_{r\mathrm{min}}, n_{r\mathrm{max}})$, where $n_{r\mathrm{min}}$ and $n_{r\mathrm{max}}$ represent the lower and upper bounds, respectively [see schematic in Fig.~\ref{fig:1}(a)]. The derivation of $n_r$ proceeds as follows.}
For a given electrostatic mode characterized by complex frequency $\omega = \omega_r + i\gamma$ and wave vector $\vec{k}$, the range $n_r$ corresponds to the interval of electron density over which this mode can serve as an unstable TPD daughter wave (i.e., with $\gamma > \nu_{ei}$, where $\nu_{ei}$ is the electron-ion collisional damping rate). 
This range can be derived from the TPD dispersion relation for a homogeneous plasma \cite{Kruer2003}
, and typical calculation results are shown in \textcolor{black}{Fig.~\ref{fig:1}(b)} for $\vec{k} = \vec{k}_a = (k_{ax} = 0.88 \omega_0/c, k_{ay} = 0.091 \omega_0/c)$, electron temperature $T_e = 2.5$~keV, pump laser intensity $I_0 = 6 \times 10^{14}$~W/cm$^2$, and wavelength \textcolor{black}{$\lambda_0 = 0.351\mu$m}. 
The selected wave vector $\vec{k}_a$ corresponds to the most unstable mode predicted by linear absolute TPD theory \cite{Simon1983} for a density scale length $L_n = 150\mu$m. The results show that $\gamma > \nu_{ei}$ within the density range $n_r = (0.2421n_c, 0.2464n_c)$, giving a resonance width $\Delta n_r \approx 0.0043 n_c$.


Although the above calculation assumes a homogeneous plasma, the resulting $n_r$ accurately reflects the spatial growth region of absolute TPD modes in inhomogeneous plasmas. Using the linear TPD simulation code LTS \cite{Yan2010}—an electrostatic fluid code benchmarked against multiple theoretical models—we compute the spatial envelope of the electron density perturbation $\delta n_e$ associated with the dominant absolute TPD mode at wave vector \textcolor{black}{$\vec{k}_a$} [\textcolor{black}{Fig.~\ref{fig:1}(c)}]. The $\delta n_e$ envelopes at two simulation times, shown with adjusted vertical scales for visual clarity, exhibit identical shapes, indicating that the mode shape (not amplitude) has reached a steady-state.
The spatial growth region of this mode is defined as the region between the peak and valley of the second derivative of the $\delta n_e$ envelope, \textcolor{black}{marked by blue and green dashed lines in \textcolor{black}{Fig.~\ref{fig:1}(c)}.} This region aligns well with the calculated $n_r$ range. 
This agreement holds across a wide range of laser and plasma conditions, as illustrated in Fig.~\ref{fig:1}(d), where the simulated growth regions match the analytically predicted $n_r$ in all cases. The underlying physical interpretation is straightforward: a mode can grow only within the regions where it is linearly unstable.

This concise physical picture naturally extends to the nonlinear saturation of TPD. 
According to the insights of $n_r$ discussed above, the growth of absolute TPD modes is disrupted—and thus saturation is reached—when density fluctuations become large enough to shift local plasma conditions outside the resonance range $n_r$ [Fig.~\ref{fig:1}(a)]. Specifically, saturation occurs when the fluctuation amplitude approaches the width of the resonance range, with a typical threshold given by \textcolor{black}{$\Delta n_{\mathrm{r}}/2$}. \textcolor{black}{Theoretical calculations show that $\Delta n_r$ scales primarily with $I^{1/2}$ while remaining insensitive to $T_e$[Fig.~\ref{fig:1}(e)]}.


The proposed saturation criterion is validated through first-principle, fully kinetic particle-in-cell (PIC) simulations. A series of simulations are performed using the OSIRIS code \cite{Fonseca2002}, under laser-plasma conditions representative of direct-drive ICF: density scale lengths $L_n \sim (150$–$300)\mu\mathrm{m}$, $T_e \sim (1$–$3)\mathrm{keV}$, and laser intensities $I_0 \sim (6$–$12)\times 10^{14}\mathrm{W/cm}^2$, as summarized in Table~\ref{tab:simu_PIC}. 
To demonstrate the generality of our findings, we include simulations with carbon (C), hydrogen (H), and mixed (CH) ion species. The electron density range is chosen as $ (0.20$–$0.27) n_c$.


\begin{table}
\caption{\protect\label{tab:simu_PIC}  PIC simulation parameters. Here $L_n$, $T_e$ and $I_0$  are initial conditions  while $I_q$ and $T_{e,eff}$ are the laser intensity at $n_c/4$ and electron temperature at saturation stage. TPD threshold factor $\eta = L_nI_0\lambda_0/(81.86T_{e}$).}
\begin{ruledtabular}
\begin{tabular}{cccccccc}
 &ion & $L_n$ &$T_e$  &$I_0$ & $T_{e,eff}$ &$I_q$ & $\eta$\\
 &&$\mu$m&keV&$ 10^{14}W/cm^2$&keV&$ 10^{14}W/cm^2$&\\
\hline
(A)&CH& 150 & 3.0  &11.0& 3.7&6.9&2.4\\
(B)&CH& 150 & 2.0  &6.0& 2.6& 4.2&1.9\\
(C)&CH& 150 & 2.0  &11.0& 2.7& 6.6&3.5\\
(D)&CH& 150 & 1.0  &11.0& 1.9& 6.4&7.1\\
(E)&CH& 150 & 3.0  &12.0& 3.8& 7.6&2.6\\
(F)&CH& 300 & 3.0  &6.0& 3.3& 2.7&2.6\\
(G)&CH& 300 & 2.0  &6.0& 2.5& 2.6&3.9\\
(H)&CH& 200 & 3.0  &9.0& 3.5& 4.9&2.6\\
(I)&CH& 200 & 2.0  &6.0& 2.5& 3.5&2.6\\
(J)&H& 150 & 3.0  &12.0& 3.6& 6.7&2.6\\
(K)&C& 300 & 3.0  &6.0& 3.4& 3.0&2.6\\
\end{tabular}
\end{ruledtabular}
\end{table}

\begin{figure}
\includegraphics[width=1.0\linewidth]{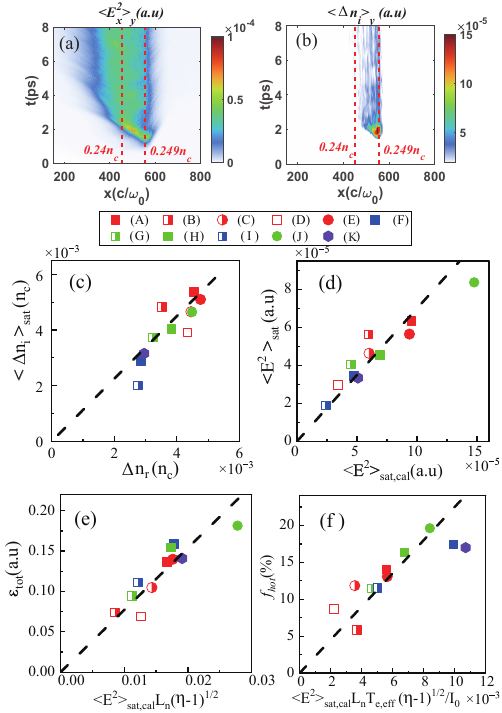}%
\caption{\label{fig:theory_simu_compare} The results of PIC simulations in table (\ref{tab:simu_PIC}). 
Spatiotemporal evolution of (a) the $E_x$ field energy and (b) ion density fluctuations $\Delta n_i$ from \textcolor{black}{simulation (B)}. 
(c) Saturation amplitude of ion density fluctuations $\langle \Delta n_i \rangle_\mathrm{sat}$ from simulations versus the theoretical $\Delta n_r$. 
(d) Comparison between theoretically predicted $\langle E^2 \rangle_{\mathrm{sat,cal}}$ and simulated $\langle E^2 \rangle_{\mathrm{sat}}$. 
(e) Simulated total TPD energy $\varepsilon_\mathrm{tot}$ versus scaling formula using Eq.~(\ref{eqn:Etot}). 
(f) Simulated $f_\mathrm{hot}$ versus the scaling model from Eq.~(\ref{eqn:fhot}). The dashed lines in (c)-(f) denote linear fittings.
}
\end{figure}

In these simulations, the growth and saturation of TPD modes are characterized by the spatiotemporal evolution of $\langle E_x^2 \rangle_y(x,t)$, where $E_x$ is the electric field component along the $x$-direction (laser incidence direction) —directly associated with the energy of TPD modes—and $\langle \cdot \rangle_y$ denotes spatial averaging over the $y$-direction.
\textcolor{black}{Fig.~\ref{fig:theory_simu_compare}(a)} shows a representative evolution of $\langle E_x^2 \rangle_y(x,t)$ from \textcolor{black}{simulation (B) }in Table~\ref{tab:simu_PIC}.
The fields initially grow near the $n_c/4$ region, 
then propagates toward lower density, and finally reaches a quasi-steady saturated state—consistent with previous findings \cite{Yan2012,Zhang2014}.
Consequently, the driven ion density fluctuation $\langle \Delta n_i \rangle_y(x,t)$ exhibits similar growth and saturation behavior, as shown in \textcolor{black}{Fig.~\ref{fig:theory_simu_compare}(b)}. 
The saturation levels of ion density flucatuations 
\textcolor{black}{denoted as 
$\langle \Delta n_i \rangle_\mathrm{sat}$ }are determined as the standard deviations near $n_c/4$ region
 during the quasi-steady phase ($t > 4\,\mathrm{ps}$), 
and the results are compared with the theoretical $\Delta n_r$ in \textcolor{black}{Fig.~\ref{fig:theory_simu_compare}(c)}. 
To ensure accuracy, all calculations of $\Delta n_r$ in this work use the effective steady-state temperature $T_{e,eff}$ as the electron temperature and intensity $I_q$ at $n_c/4$ as the laser intensity. 
The results show a consistent scaling of \textcolor{black}{$\langle \Delta n_i \rangle_\mathrm{sat} \approx 1.1\Delta n_r$} across a wide range of laser and plasma conditions, indicating that \textcolor{black}{$\Delta n_r$} robustly characterizes the fluctuation amplitude at TPD saturation.

The saturated electric field amplitude 
in the quasi-steady state can be estimated by balancing the ponderomotive force $F_p= -\frac{e^2}{4m_e \omega_{pe}^2} \nabla \langle E^2 \rangle_{te}$, where $\langle \cdot \rangle_{te}$ denotes averaging over timescales longer than the electron plasma wave period but shorter than ion response times, with the gradient of low-frequency plasma pressure perturbations:
$F_p = -\frac{1}{n_e} \nabla \left( \tilde{P}_{el} + \tilde{P}_{il} \right)
$, 
where $\tilde{P}_{el} = \tilde{n}_{el} T_e$ and $\tilde{P}_{il} = 3 \tilde{n}_{il} T_i$ represent the low-frequency pressure perturbations of electrons and ions, respectively, under isothermal (electrons) and adiabatic (ions) assumptions. 
The corresponding density perturbations $\tilde{n}_{el}$ and $\tilde{n}_{il}$ are associated with the observed ion density fluctuations. 
Then we have
\begin{align}
\label{eqn:force_balance_2}
<E^2>_{te} \sim 16\pi(ZT_e + 3T_i) \tilde{n}_{il}
\end{align}
where $Z$ refers to the ion charge state. The above \textcolor{black}{Eq.~}(\ref{eqn:force_balance_2}) omits a possible constant of integration arising from the removal of the gradient operator, and thus does not represent an exact relation between the ion density fluctuations and the turbulent electric fields. Nevertheless, it can serve as a useful approximation. 
By performing spatial ($x$, $y$) and temporal ($ti$) averages—where $ti$ exceeds the ion response time—on both sides of Eq.~(\ref{eqn:force_balance_2}), we obtain the calculated saturation amplitude of electric fields:
\begin{align}
\label{eqn:force_balance_3}
\langle E^2 \rangle_{\mathrm{sat,cal}} ~\sim ~ 16\pi(ZT_e + 3T_i) \langle \Delta n_i \rangle_\mathrm{sat}
\end{align}
Fig.~\ref{fig:theory_simu_compare}(d) shows the comparison between the simulated $\langle E^2 \rangle_{\mathrm{sat}}$ and \textcolor{black}{the calcualted values $\langle E^2 \rangle_{\mathrm{sat,cal}}$ derived from the simulated \textcolor{black}{$\langle \Delta n_i \rangle_\mathrm{sat}$} using Eq.~(\ref{eqn:force_balance_3})}. Despite the simplifications involved, the two remain approximately proportional, with a factor of \textcolor{black}{$\sim$0.7}. 

With the saturated electric field amplitude, the hot electron energy can be reasonably evaluated. 
Directly calculation 
is challenging due to the turbulent waves with broad spectrum. 
However, since the turbulent wave fields originate from TPD,
their spectral characteristics should retain a degree of similarity across different conditions. 
This allows the total Langmuir wave energy, \( \varepsilon_{{tot}} \), to serve as a proxy for estimating hot electron energy. 
We find that \( \varepsilon_{{tot}} \) scales with 
$\langle E^2 \rangle_{\mathrm{sat,cal}}$ and the effective spatial range \( S_{{sat}} \) over which TPD modes grow. Theoretically, \( S_{{sat}} \) increases monotonically with both the TPD threshold factor \( \eta \) and the density scale length \( L_n \). The simulation results demonstrate a fair empirical relation:
\begin{align}
\label{eqn:Etot}
\textcolor{black}{\varepsilon_{tot}} \sim \langle E^2 \rangle_{\mathrm{sat,cal}} L_n (\eta - 1)^{1/2},
\end{align}
as shown in Fig.~\ref{fig:theory_simu_compare}(e). Here the factor $\eta - 1$ indicates that TPD growth occurs exclusively when $\eta > 1$. 

Besides, the ability of the Langmuir field to accelerate electrons also depends on the electron temperature $T_{e,eff}$, which influnces the number of background electrons that can be accelerated by Langmuir waves. 
Incorporating this effect, we find that $f_{hot}$ across various laser-plasma conditions is well characterized by
\begin{align}
\label{eqn:fhot}
f_{{hot}} \sim \langle E^2 \rangle_{\mathrm{sat,cal}} L_n (\eta - 1)^{1/2} T_{e,eff}/I,
\end{align}
as demonstrated in Fig.~\ref{fig:theory_simu_compare}(f),
suggesting the validity of the Eq. (\ref{eqn:fhot}) in describing $f_{hot}$. 


The trend of actual $f_{hot}$ for varying $I$ \textcolor{black}{can be estimated} using the above \textcolor{black}{Eq.~(\ref{eqn:fhot})}. As we find $\Delta n_r\sim I^{1/2}$ from  Fig.~\ref{fig:1}(e) and notice $L_n\propto I^{1/3}$ and $T_{e,eff}\propto I^{2/3}$ from conventional 1-D ablation model\cite{Atzeni2004}, 
we have \textcolor{black}{$\langle E^2 \rangle_{\mathrm{sat,cal}} \sim T_{e,eff}\Delta n_r\sim I^{7/6}$ }[\textcolor{black}{Eq. (\ref{eqn:force_balance_3})}]
 and 
\begin{align}
\label{eqn:fhot_I}
f_{{hot}} \approx \alpha I^{7/6} (\beta I^{2/3} - 1)^{1/2},
\end{align}
where we consider $\eta\sim IL_n/T_e\sim I^{2/3}$. 
The coefficients $\alpha$ and $\beta$ relate $I$ to $\langle \Delta n_i \rangle_\mathrm{sat}$ and $\langle E^2 \rangle_\mathrm{sat}$, ultimately connecting to $f_\mathrm{hot}$, where they encode the underlying physical details. 
For instance, laser beams with identical overlapped intensity $I$ but different geometries should produce distinct relationships between $I$ and $\langle \Delta n_i \rangle_\mathrm{sat}$, while the same total electric field energy $\varepsilon_\mathrm{tot}$ distributed for different TPD mode spectra will generate varying $f_\mathrm{hot}$ values.
Therefore, $\alpha$ and $\beta$ vary with laser plasma conditions and configurations, while the scaling formula Eq.~(\ref{eqn:fhot_I}) can be reasonably maintained. 

\begin{figure}
\includegraphics[width=0.90\linewidth]{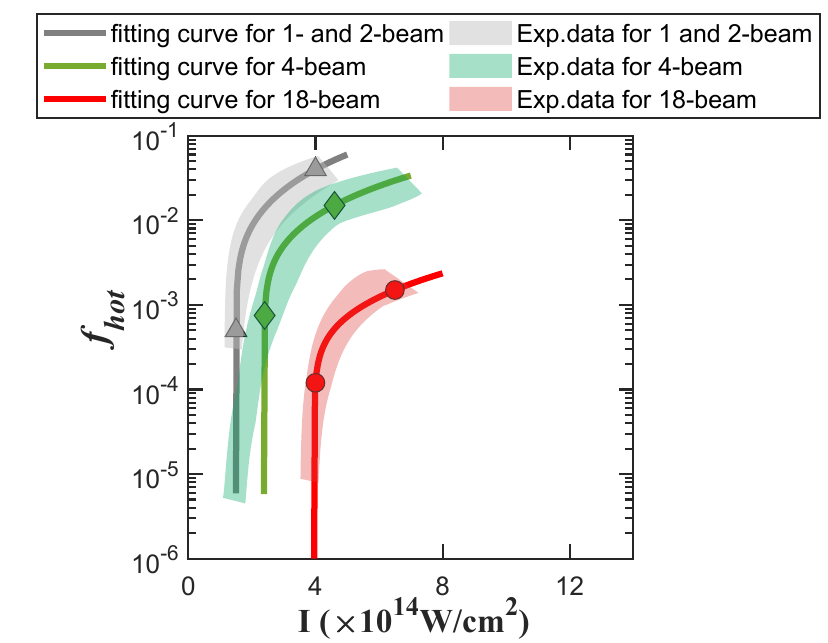}%
\caption{\label{fig:EXP} \textcolor{black}{
Calculated relationships between \( f_{\mathrm{hot}} \) and \( I \) for three laser beam configurations: 1- and 2-beam (gray), 4-beam (green) and 18-beam (red). Each curve follows Eq.~\eqref{eqn:fhot_I} with $\alpha$ and $\beta$ calibrated with two experimental data points extracted from Ref.~\cite{Michel2013}. The shaded areas highlight the regions of concentrated experimental data points for each beam configuration. For 1- and 2-beam case, $\alpha = 8.2\times 10^{-3}, \beta = 0.8$; for 4-beam case, $\alpha = 3.4\times 10^{-3}, \beta = 0.6$; and for 18-beam case , $\alpha = 2.7\times 10^{-4}, \beta = 0.4$.}
}
\end{figure}

This scaling formula solely depends on laser intensity $I$ and requires only two data points—pairs of $(I, f_{hot})$—to determine $\alpha$ and $\beta$ for certain laser plasma conditions.
\textcolor{black}{As \( f_{\mathrm{hot}} \) values from experiments and small-scale simulations have been shown to agree when the latter sufficiently sample laser-plasma conditions~\cite{Follett2017}, the scaling relation in Eq.~(\ref{eqn:fhot_I}) is expected to reflect experimental measurements.}
Through the application of this methodology, we successfully reproduce the experimentally observed relationship between $f_{hot}$ and laser intensity across multiple laser configurations at OMEGA and OMEGA-EP facilities\cite{Michel2013}, as illustrated in Fig.~\ref{fig:EXP}. Using two initial data points (represented by colored markers roughly extracted from Fig.~2 of reference \cite{Michel2013}), we obtain the corresponding fitting curves of $f_{hot}$ for three different laser beam configurations. These curves demonstrate excellent agreement with the remaining experimental data from the same setups. 
The strong consistency between theory and experiment underscores the robustness of the proposed scaling Eq.~(\ref{eqn:fhot_I}).


In summary, 
we provide new insights into TPD by identifying the resonance density range $\Delta n_r \sim I^{1/2}$ of absolute modes as the key determinant of nonlinear saturation amplitudes of ion density fluctuations and Langmuir waves. 
This persists despite the presence of a broad TPD spectrum extending beyond conventional absolute modes and the interplay of multiple saturation mechanisms.
From the saturated Langmuir wave amplitudes, we derive a predictive scaling formula for the hot-electron energy fraction $f_{\text{hot}}$ as a function of laser intensity $I$. 
When calibrated with just two experimental $(I, f_{\text{hot}})$ data points, this formula can reproduce the observed $f_{hot}$-$I$ relation on OMEGA and OMEGA EP facilities. 
This work establishes a potential reliable tool to predict TPD-driven hot electrons for ICF experiments.

\begin{acknowledgments}
This research was supported by the National Natural Science Foundation of China (NSFC) (Grant Nos. 12275269, 12388101, 12375243 and 12275032), by the Strategic Priority Research Program of Chinese Academy of Sciences (Grant Nos. XDA25010200 and XDA25050400). The numerical simulations in this paper were conducted on Hefei advanced computing center.
\end{acknowledgments}

\appendix




\bibliography{apssamp}

\end{document}